\documentclass[aps,prl,reprint,superscriptaddress]{revtex4-1}
\usepackage[T1]{fontenc}
\usepackage{amsmath}
\usepackage{amssymb}
\usepackage{mathalfa}
\usepackage{hyperref}

\usepackage{cases}
\usepackage{verbatim}
\usepackage{amsthm}
\usepackage{subcaption}
\usepackage{xcolor}
\usepackage{enumitem} 
\usepackage{mathrsfs}
\usepackage{graphicx,wrapfig,lipsum} % Required for inserting images

\begin{document}

\title{Study of the Emergence of a Gluon Mass Scale from Center Vortices Using a Wave-Functional Formalism}
\author{David R. Junior}
\affiliation{
Instituto de F\'isica Te\'orica, Universidade Estadual Paulista and South-American Institute of Fundamental Research ICTP-SAIFR, Rua Dr. Bento Teobaldo Ferraz, 271 - Bloco II, 01140-070 S\~ao Paulo, SP, Brazil}
\affiliation{Institut f\"ur Theoretische Physik, Universit\"at T\"ubingen, Auf der Morgenstelle 14, 72076 T\"ubingen, Germany}
\author{Gast\~ao  Krein}

\affiliation{
Instituto de F\'isica Te\'orica, Universidade Estadual Paulista and South-American Institute of Fundamental Research ICTP-SAIFR, Rua Dr. Bento Teobaldo Ferraz, 271 - Bloco II, 01140-070 S\~ao Paulo, SP, Brazil}
\author{Luis E. Oxman}
\affiliation{
Instituto de F\'isica, Universidade Federal Fluminense, 24210-346 Niter\'oi, RJ, Brasil.} 
\author{Bruno R. Soares}

\affiliation{
Instituto de F\'isica Te\'orica, Universidade Estadual Paulista and South-American Institute of Fundamental Research ICTP-SAIFR, Rua Dr. Bento Teobaldo Ferraz, 271 - Bloco II, 01140-070 S\~ao Paulo, SP, Brazil}
\date{\today}
\begin{abstract}
Lattice simulations and theoretical analyses consistently identify center vortices and monopoles as key nonperturbative configurations in Yang-Mills theory. In the continuum, the effective representation of mixed oriented and nonoriented center vortices showed that these degrees of freedom generate a confining flux tube with $N$-ality. Independently, studies of correlation functions reveal an infrared behavior characterized by massivelike scales. In this Letter, field correlators are computed for the first time in a theoretical framework based on center vortices. Using an Abelian-projected vacuum wave functional peaked on the mixed ensemble, we show the emergence of a massivelike gauge-invariant field strength correlator. For this behavior, the nonoriented component in the center-vortex condensate turns out to be essential, as is also the case for producing the correct properties of confining flux tubes.

\end{abstract}

\maketitle

\section*{Introduction}

The confining nature of pure Yang-Mills (YM) theory has long been established through Monte Carlo lattice simulations, which show that the Wilson loop for fundamental quark probes obeys an area law at large distances \cite{qq1,qq2}. Since then, a continuous effort has been made to advance an understanding of the mechanism underlying the infrared (IR) properties of non-Abelian gauge theories, employing lattice methods and formulations in continuum spacetime. 

Gauge field correlation functions in the continuum have been studied based on the refined Gribov-Zwanziger formalism~\cite{gribov1}, the Dyson-Schwinger equations~\cite{cristina,benn}, and functional renormalization group methods~\cite{frg}. All these approaches have produced IR two-point correlation functions displaying a dynamically generated gluon mass scale, as originally proposed in Ref.~\cite{Cornwall:1981zr} and in agreement with the behavior observed in lattice simulations~\cite{lattprop1,lattprop2,gribov-latt,lattprop3,lattprop4,lattprop5}. Motivated by such results, effective models that incorporate a gluon mass term were proposed; see the review in~\cite{Pelaez:2021tpq} for references. In this theoretical setting, gluon confinement is associated with infrared suppression and spectral positivity violation of the correlation functions~\cite{Cornwall:2013zra}. However, a connection to confinement in the Wilson sense continues to be missing when dealing with these gauge-dependent correlators. Theoretical studies in the maximal Abelian gauge (MAG) \cite{MAG-G} and the large-distance behavior of field-strength correlators in the Landau gauge extracted from lattice simulations \cite{landau-g} also reveal the emergence of infrared mass scales.

On the other hand, gauge-invariant correlators, investigated only by using lattice methods, do display a connection with the Wilson criterion. In particular, two-point correlators of field strengths connected by holonomies have also revealed the emergence of a nonperturbative mass scale \cite{ref:gaugeinv-su2,ref:gaugeinv, ref:gaugeinv-1, ref:gaugeinv-2}, in the form of an exponential falloff. The large distance properties of these correlators have been extensively studied in the lattice. Building on this physical input, complemented by a stochastic model of vacuum fluctuations, a nonvanishing string tension was obtained \cite{Dosch:1987,Simonov:1988,
DiGiacomo:2002}.

Another important approach is based on early groundbreaking proposals based on topological configurations such as center vortices and monopoles \cite{Mandelstam1976,tHooft1978,tHooft1979,MackPetkova1979,NielsenOlesen1979}. Along these lines, essential contributions using lattice methods have been made over the years \cite{Kronfeld1987,Bali1996,DelDebbio1997,DelDebbio1998,Forcrand1999,Langfeld1999,Alexandrou2000,Ambjorn2000,randomsu2,Forcrand2001,Faber2001,Koma2003,randomsu3,bowman2011,omalley2012,Hollwieser2013,Stokes2014,Sakumichi2014,biddle2020,mickley2024}. 
Sophisticated detection techniques have revealed the presence of these objects in Monte Carlo configurations, allowing for the determination of their properties and their relevance to the Wilson loop area law at large distances. Center vortex degrees of freedom play a fundamental role in the observed $N$-ality properties of the Wilson loop at asymptotically large distances \cite{n-ality,double-wl}. Here, $N$-ality refers to the dependence of the asymptotic string tension solely on how the center Z(N) of the gauge group SU(N) is realized in the quark representation. A relationship has also been established between the IR behavior of correlation functions and center vortices \cite{Biddle2018}. More specifically, lattice simulations indicate that smoothed center vortices are connected with the infrared suppression observed in the gluon propagator, computed in the Landau gauge. An understanding of this phenomenon from the usual first-principles theoretical setting is hard, as it considers topological configurations only in connection with gauge fields near the Gribov horizon, where the perturbative Faddeev–Popov quantization breaks down, and not as explicit elements of the formalism. See, however, Refs. \cite{Oxman2015,Fiorentini2020,Fiorentini2021,Fiorentini2022}, where a renormalizable first-principles definition of the Yang-Mills ensemble was introduced in the continuum. 

In this Letter, we present for the first time, within a well-defined theoretical setting, a connection between a vacuum governed by oriented and nonoriented center vortices, on the one hand, and field correlators displaying a gluon mass scale, on the other. More precisely, we will assess the deep-IR properties of gauge-invariant field-strength correlators in momentum space. 

In Refs.~\cite{Ambjorn2000,Forcrand2001}, center vortices and Cartan monopoles were found to be interconnected in the YM vacuum, forming nonoriented center vortices or chains. In SU(2), the flux emanating from monopoles, identified in Abelian-projected configurations, was shown to be collimated along pairs of center vortices. Similar measurements in SU(3) are consistent with this picture~\cite{Stack2002}. Furthermore, the combination of oriented and nonoriented center vortices can carry a nonvanishing topological charge~\cite{Engelhardt2000}.

A theoretical framework in which an ensemble of collimated oriented and nonoriented center vortices induce a confining electric flux tube between quarks, and thus a Wilson loop area law, was proposed in Ref.~\cite{mixed}. The monopoles are non-Abelian, with a magnetic flux that is not isotropically distributed. Instead, they interpolate a pair of center vortices carrying different Cartan fluxes (nonoriented center vortices). For a detailed discussion and visualization of these configurations, although in the Higgs phase, see Ref. \cite{conf-qg}. In this proposal, it was argued that the partition function for percolating center-vortex world surfaces can be described by non-Abelian gauge-field Goldstone modes, while monopole worldlines are captured by effective adjoint scalar Higgs fields. This yields the appropriate field content for the spontaneous symmetry-breaking pattern SU(N) $\to$ Z(N), which supports confining flux tubes with the correct asymptotic $N$-ality and allows for an Abelian-like transverse action density (see \cite{prospecting} and references therein). 
This ensemble of oriented and nonoriented center vortices, constructed in terms of elementary vortex branches that can be matched in groups of $N$, was reassessed in Ref.~\cite{wavefunctional}. In that work, a wave functional for the Abelian-projected configurations peaked on the mixed ensemble was proposed. Recently, these ideas were settled in the lattice,  based on a matrix-model representation of  surfaces~\cite{weingarten-new}. In this way, the relation between percolating world surfaces and gauge-field Goldstone modes was firmly established. These works clarified why an Abelian-projected scenario can display confining flux tubes with $N$-ality, as observed in the lattice \cite{Bali1996,Ambjorn2000,Koma2003,Sakumichi2014}. This scenario was reviewed in Refs. \cite{review-paper,proc}, which include a visualization of possible center-vortex matchings.

Here, the connection between center vortices and field correlators will be implemented within the wave functional formalism. This description of the mixed ensemble allows one to trace the probability amplitude of locally Abelian gauge-field variables, giving access to gauge-invariant field-strength correlators. On the one hand, our results help bridge two traditionally separate communities: those focused on computing correlators and those studying center-vortex ensembles. On the other hand, this adds to the observed confining properties previously accommodated at asymptotically large distances (for a recent status, see \cite{weingarten-new}): confining flux tubes with $N$-ality, difference-of-areas law for double Wilson loops in SU(2) and Y-shaped potentials for SU(3), the asymptotic Casimir law, Abelian-like profiles, and possible deviations, thus supporting the mixed-ensemble scenario as a robust candidate for the confinement mechanism in Yang-Mills theory.
 \\
\indent Before deriving the field correlators associated with the mixed ensemble, we emphasize that the theoretical framework for confinement developed in Refs. \cite{mixed,prospecting,wavefunctional,weingarten-new} is phenomenological in origin. It is based on identifying the different elements observed in lattice studies, singling out the essential ones, and deriving an effective description for them. The proposal relies on percolating elementary center vortices carrying Cartan fluxes, their natural correlations, and the possibility of nonorientation. As in other areas of physics, its validation is assessed by comparing its predictions with independent observations.\\
\indent In a more theoretical context, it is interesting to note recent advances in theories with $Z_N$ one-form symmetry, where ensembles only formed by percolating center-vortex loops, with no correlations among them, were shown to display a Coulomb nonconfining phase \cite{unsal}. Also, regarding a first-principles calculation, we would like to comment that the YM ensemble of Refs. \cite{Oxman2015,Fiorentini2020,Fiorentini2021,Fiorentini2022} is ultimately originated from Singer's no-go theorem~\cite{singer-0}. While this theorem forbids a global gauge fixing, it does not forbid a partition of the configuration space into sectors where the usual Faddeev-Popov procedure can be locally performed. Thus, inspired by the lattice Laplacian center gauge~\cite{Faber2001,Forcrand2001}, these sectors were implemented by equivalence classes of singular mappings $S$, associated with oriented and nonoriented center vortices. Despite this progress, no systematic approximation framework has yet been developed to characterize their dominant features. This represents a challenging open problem that lies beyond the scope of the present work. Interestingly, nonoriented center vortices were recently shown to be a preferred saddle point for YM theory in compactified spacetimes \cite{yuya}. 

\section{Gauge-invariant correlators  due to center vortices}
\label{review}

 Let us consider the gauge-invariant object
\begin{align}
 O_F(C_1, C_2) = {\rm Tr}\{F_{\mu \nu}(x)\Gamma_{C_1}[A]F_{\rho \sigma}(y)\Gamma_{C_2}[A]\}\;,\label{def-gi}
\end{align}
where $F_{\mu\nu}$ is the gauge-covariant field strength, $\Gamma_C$ is an holonomy, and $C_1, C_2$ are arbitrary paths connecting $x$ to $y$ and $y$ to $x$, respectively (the spacetime indices in $O_F$ are understood). On the lattice, it was studied by taking a line $C$ and the oppositely oriented $-C$. In this work, we shall compute them by assuming Abelian dominance, understood as due to the dominance of both oriented and nonoriented center vortices. These are non-Abelian fields that are locally Abelian. In this respect, consider the subspace of the configuration space of Yang-Mills fields, parametrized by $\bar{\mathcal{A}}$ and the map $S\in$ SU(N) \cite{cho,cho-sun}: 
\begin{align}
A&=\bar{\mathcal{A}}^q n_q+i [n_q,\nabla n_q] \;, \label{g-A} \\ 
n_q &= ST_q S^{-1} \;,\label{our-a}
\end{align}
$T_q$, $q=1, \dots, N-1$, are the Cartan generators of $\mathfrak{su}(N)$. The associated field strength, $F_{\mu \nu}=G_{\mu \nu}^q n_q$, is
\begin{align}
  G^q_{\mu\nu}=(\partial_\mu\bar{\mathcal{A}}^q_\nu-\partial_\nu\bar{\mathcal{A}}^q_\mu+H_{\mu\nu}
    ^q)\;,
    \label{Fst}  
\end{align}
where $H^q_{\mu\nu}$ encodes the monopole current density. For example, in SU(2) there is just one diagonal generator $T_1 \propto \sigma_3$. In this case, the projection of the chromomagnetic field $B_i =\frac{1}{2}\epsilon_{ijk} F_{jk} $ along the local $\mathfrak{su}(2)$ direction $ n_1= S T_1 S^{-1} $ yields
\begin{align}
    \left(\nabla \times \bar{\mathcal{A}}\right)_i +  \frac{\epsilon_{ijk}}{2} \hat{n} \cdot (\partial_j \hat{n} 
 \times \partial_k \hat{n} ) 
 \; ,
 \label{Adj-su2}
\end{align}
where $\hat{n}$ lives on the unit sphere $S^2$, and $S \sigma_3 S^{-1} = \hat{n} \cdot \vec{\sigma}$. 
The first term is an Abelian-like magnetic field, while the second is proportional to the topological charge density. Indeed, when integrated over a closed surface $\mathcal{M}$, it gives the monopole charge as $4\pi$ times the topological charge of
the mapping $\mathcal{M} \to S^2$.

As shown in Ref. \cite{Abe-Dom} for 
SU(2), and later generalized to SU(N) in Ref. \cite{conf-qg}, this  subspace includes not only oriented but also nonoriented center vortices. For example, on a fixed-time slice, a  monopole in SU(2) is originated at a pointlike defect in $\hat{n}$. When $\hat{n}$ is orthogonal to a surface $S^2$ around this point, the directions $\hat{n}_2$ and $\hat{n}_3$, which are given by the adjoint rotation in Eq. \eqref{Adj-su2}, but applied to the off diagonal Pauli matrices, are necessarily tangent to $S^2$. Therefore they must have defects. Usually, this is associated to a single unobservable Dirac string, where  the local frame components $\hat{n}_2$, $\hat{n}_3$ rotate twice when going around. However, there could also be a pair of observable lines, emanating from the monopole, where the frame rotates once. As is well known from the spin-$1/2$ representation, this behavior in $R \in$ SO(3) corresponds to a sign change in $S \in$ SU(2) when going around the loop. The generation of this center-element is a defining property of center vortices. In the situation we just described, the pair is correlated with a monopole, forming a nonoriented object in the Lie algebra. 

For general $N$, introducing the field $C_\mu = C_\mu(R)$,
\begin{align}
\mathcal{R} (C_\mu) =i R^{-1}\partial_\mu R \;,
\end{align}
where $R=\mathcal{R}(S)$ is in the adjoint representation of SU(N), and the subspace in Eq. \eqref{our-a} is obtained from
\cite{conf-qg}
\begin{align}
    \mathcal{R} (A_\mu) &= R \mathcal{R}(\mathcal{A}_\mu) R^{-1}+iR\partial_\mu R^{-1}\;, \label{Ad-g} \\
\mathcal{A}_\mu &= (\bar{\mathcal{A}}^q_\mu +C^q_\mu)\, T_q  \;.
\end{align}
 Again, when $R(S)$ contains defects but is single valued when going around any loop, $S$ may change by a center element $e^{\pm i\frac{2\pi k}{N}}$, characterizing the presence of center vortices. In the projections of $B_i$
 along $n_q$
\begin{align}
     G^q &= (n^q ,B) \nonumber \\ &= \nabla\times\bar{\mathcal{A}}^q + H^q \;, 
    \label{Gqi}
\end{align}
and the monopole charge densities are more easily visualized in the combinations
\begin{align}
    &\beta|_q H^{q}_i = i\frac{\epsilon_{ijk}}{2} (P_\beta,[\partial_j P_\beta,\partial_k P_\beta])\;, 
\end{align}  
 where $P_\beta =\beta|_q n_q$ and the tuple $\beta$ is a magnetic weight of the defining representation.
 
 On the locally Abelian subspace, which is gauge invariant, the observable in Eq. \eqref{def-gi} gets simplified when using $C$ and $-C$. Unlike the fundamental Wilson loop, it is blind to the center, as the holonomies appear
 in the combination ${\rm Tr}\, (T_A\Gamma T_B\Gamma^{-1}) = \mathcal{R}(\Gamma)|_{AB}$. Indeed, from Eq. \eqref{Fst} and $n_q = T_A R_{A q}$,
 \begin{align}
     {\rm Tr} [F(x)\Gamma F(y) \Gamma^{-1}]  =G^q(x)G^q(y)  \;,
 \end{align}  
 where we used $\mathcal{R}(\Gamma[A])= R(x) \mathcal{R}(\Gamma[\mathcal{A}]) R^{-1}(y)$
 [cf. Eq. \eqref{Ad-g}]. That is, the
 electric and magnetic correlators $O_E(C,-C)$ and $O_B(C,-C)$, which have the same form as in Eq. \eqref{def-gi}, with $E_i$ and $B_i$ in the place of $F_{\mu \nu}$, become local:
\begin{align}
O_E(x,y) &=G_{0 i}^q(x)G_{0 j}^q(y), \\ O_B(x,y) &=G_{i}^q(x)G_{j}^q(y)\;.
\end{align}

 Here, we shall study these observables in the Schr\"odinger picture, where the vacuum is described by a wave functional ($\Psi$) that depends on the gauge-field spatial components on a given time slice $\mathbb{R}^3$. 
 The main gauge-invariant information about collimated fluxes, which are localized around $\bar{\mathcal{A}}_i =-C_i(R)$, is given by their guiding centers $\gamma$,  where $R$ contains defects, as well as the distribution of defining magnetic weights $\beta_i$, $i=1, \dots, N$ of elementary vortices. For typical configurations, we have (see the Supplemental Material \cite{suppl-mat}).
 \begin{align}
  -C(R)=   a(\{\gamma\})= \sum_{\{\gamma\}}a_\beta(\gamma) \;.
\end{align}
The networks considered $\{\gamma \}$ include elementary loops, arrays with $N$ matching where $N$ vortex lines meet at a common point, and chains which are formed by a set of vortex lines carrying different charges, interpolated by monopoles, forming a configuration with a flux concentrated on the vortex lines. These configurations were chosen in alignment with observations in the lattice. In this respect, large clusters of percolating center vortices were observed in Refs. \cite{percolating-1,percolating-2}. For SU(3), the matching of three vortex lines was observed in Refs.   \cite{branching,mickley2024}, and collimated chains were observed for SU(2) in Ref.  \cite{Ambjorn2000}. 

In Ref.~\cite{wavefunctional}, in order to avoid irrelevant gauge-dependent information about the frame, we wrote the wave functional with variables fixed in the MAG. Indeed, there is a single-valued gauge transformation $S_{\rm D}$ such that the gauge field in Eqs. \eqref{g-A} and \eqref{Ad-g} is physically equivalent to 
\begin{align}
A_{\rm MAG}&=[\bar{\mathcal{A}}^q+C^q(R_{\rm D}) ]T_q\;,     \label{Bmag} \\
        F^{\rm MAG}_{\mu\nu} &= G_{\mu\nu}^qT_q+iS_D^{-1}[\partial_\mu,\partial_\nu]S_D \;,
        \label{comm}
\end{align}
where $F_{\rm MAG}$ is simply the Abelian field strength of $A_{\rm MAG}$. 
The commutator of ordinary derivatives (Dirac string) is also present when dealing with the non-Abelian 't Hooft-Polyakov monopole when forcing the Higgs field to be along the Cartan sector. In addition, this is consistent with Eq. \eqref{Gqi}. From Eq. \eqref{comm}, we can write
\begin{gather}
    G^q = \nabla \times \bar{\mathcal{A}}^q -\nabla \zeta^q\;,
    \label{Bmag}
\end{gather} 
where $-\nabla \zeta$ is Coulomb-like, with divergence localized at the monopoles, which is the gauge-invariant physical information about the map $n^q$. The proposed wave functional, peaked at the networks, reads as
\begin{equation}
\Psi(\bar{\mathcal{A}},\zeta) =\sum_{\{\gamma\}}\psi_{\{\gamma\}}  \delta(\bar{\mathcal{A}}-a(\{\gamma\})) \delta(\zeta-(-\Delta)^{-1}\nabla\cdot b(\{\gamma\}))\;,
\end{equation}
where $\bar{\mathcal{A}}$ is fixed with $\nabla \cdot \bar{\mathcal{A}} = 0$; this is the appropriate Coulomb condition to be used in the Hamiltonian formalism. Here, $\psi_{\{\gamma\}}$ is a weight that contains stiffness and tension, and $ b(\{\gamma\})$ is the flux density along the network.

\section{Gauge-invariant correlators}

When writing quantum states in terms of $\Psi(\bar{\mathcal{A}}, \zeta)$, the chromomagnetic and chromoelectric field operators are c numbers and functional derivatives, respectively. However, the effective theory for the quantum ensemble is known in the electric field representation, which corresponds to performing a functional Fourier transform $\Psi(\bar{\mathcal{A}},\zeta) \to \tilde{\Psi}(\mathcal{E}, \eta)$ and integrating over the ensemble. In this language, to generate translations of the magnetic field, the electric field is given by the c number operator (see the Supplemental Material)
\begin{align}
    E_{\rm MAG}= \mathcal{E} +\frac{\nabla \eta}{(-\nabla^2)^\frac{1}{2}}\;.
\end{align} 
Then, the equal-time two-point functions read as
\begin{align}
    &\langle O_E (C, -C)\rangle=\int D\mathcal{E}D\eta \, |\tilde{\Psi}(\mathcal{E},\eta)|^2 O_E (x,y)\;, \\
    &\langle O_B (C, -C)\rangle=\int D\bar{\mathcal{A}}D\zeta\, |\Psi(\bar{\mathcal{A}},\zeta)|^2 O_B(x.y)\;.
\end{align}
Of course,  for the averaged magnetic correlator in dual language, when acting on $ \tilde{\Psi}(\mathcal{E},\eta)$, $\bar{\mathcal{A}}$ and $\zeta$ in Eq. \eqref{Bmag} become functional derivatives with respect to $\mathcal{E}$ and $\eta$, respectively. This
wave functional is given by \cite{wavefunctional}
\begin{align}
        \tilde{\Psi}(\mathcal{E},\eta) =  \int [D\Phi^\dagger][D\Phi]\, e^{-W(\Phi,\Lambda)}\;, \label{wavefunctional_first}
\end{align}
where $\Phi$ is an $N\times N$ diagonal matrix, with complex field entries $\phi_i$ carrying charge $\beta_i$, minimally coupled to a dual Cartan gauge field $\Lambda = \Lambda^T + \Lambda^L$,
\begin{align}
        \Lambda^T  &= 2\pi 2N \frac{\nabla \times \mathcal{E}}{-\nabla^2}, \nonumber \\ \Lambda^L&= 2\pi 2N \frac{\nabla \eta }{-\nabla^2}\;. 
        \label{Ltl}
\end{align}
The field $\Phi$ is governed by the action ($D=\nabla - i \Lambda$)
\begin{align}
        W(\Phi,\Lambda) &= \int d^3x  \left( {\rm Tr} ((D(\Lambda)\Phi)^\dagger D(\Lambda)\Phi)+V(\Phi)\right)\;, \nonumber\\  V(\Phi) &= \frac{\lambda}{2}{\rm Tr}(\Phi^\dagger\Phi-a I_N)^2-\xi(\det \Phi+\det\Phi^\dagger)\nonumber\\ &-\vartheta {\rm Tr}(\Phi^\dagger T_A \Phi T_A)\;.
        \label{sec-or}
\end{align}
In the center-vortex condensate $a> 0$, when the nonoriented component is  turned off ($\vartheta =0$), there are massless Goldstone fluctuations $\Phi \propto S$, where $S$ is in the Cartan subgroup of SU(N). For a mild $\vartheta$, the vacua become discrete. Thus, to compute correlation functions, we consider the soft massive phase fluctuations around the vacuum $\Phi= v 1_{N}$:
\begin{align}
    \Phi \approx v e^{i\frac{\theta^pT_p}{v}}, \label{effective_scalar_fluctuations} 
\end{align} with $\theta^p(x)  \in \mathbb{R} $. Taking into account up to quadratic terms in $\theta$, the computation of observables can be done using a Gaussian path integral on the variables $\mathcal{E},\eta, \theta$. Upon integration over $\mathcal{E},\eta$, an expansion in powers of $1/v$ arises. In particular, the leading-order contribution to the correlation function of the chromoelectric field is, in momentum space, (see the Supplemental Material)
\begin{gather}
 \langle O_{E}(k) \rangle = \frac{(N-1)\pi}{4N v^2(2\pi)^3} \left[  k^2 P^T_{ij}  +  \frac{ k^2(k^2 + m^2)}{m^2} P^L_{ij}\right]\label{electricvortex}\;,
\end{gather}
with the transverse and longitudinal projectors \begin{align}
    P_{ij}^T&=\delta_{ij}-\frac{k_i k_j}{k^2}, \nonumber \\
    P_{ij}^L&=\frac{k_i k_j}{k^2},
\end{align} and $m^2 = \vartheta/2$.
The chromomagnetic correlations can also be obtained, and the result is 
 \begin{align}
  \langle O_B(k) \rangle= (2\pi 2N)^2\frac{v^2(N-1)}{2N} 
 \left[  P^T_{ij}  +  \frac{m^2}{(k^2 + m^2)} P^L_{ij}\right] \;. \label{bvortex}
  \end{align} 
  Moreover, the longitudinal component gives the magnetic charge density $\rho(k) = k_i G_i(k)$ correlation,
\begin{equation}
\langle \rho^q(k) \rho^p(-k)\rangle =\nonumber (2\pi2 N)^2 \frac{ \delta^{pq}}{2N}  \frac{v^2  m^2 k^2}{k^2+m^2} \;.
\end{equation} 
Then, we see that the mixed center-vortex ensemble generates a mass scale in the gauge-invariant correlation functions, with the nonoriented component playing an essential role in this effect.  
These results are valid only in the deep infrared, below a physical cut off $\mathcal{Q}$, which gives the validity region where the effective second-order field theory in Eq. \eqref{sec-or} can be used. This theory was obtained in an approximation in which the tangent vectors at the center-vortex endpoints are weakly correlated, in spite of stiffness. This involves distances much larger than an effective monomer size $1/\mathcal{Q}$, which acts as a UV cutoff. We also note that these results are analogous to those derived by Polyakov in Ref. \cite{polyakov} for compact electrodynamics in $2+1$ dimensions. In that case, the condensation of magnetic monopoles with Coulomb-like interactions was shown to lead to a mass gap and to confinement.
Here, it is the mixed ensemble of oriented and nonoriented center vortices in the Yang-Mills vacuum that generates the mass and drives confinement. 

The field-strength correlators have been evaluated in the lattice for SU(3) for the particular case where $C$ is a straight line connecting the points $x,y$ \cite{ref:gaugeinv}. The result is ($z\equiv |x-y|$)
\begin{align}
     \langle O_E (C, -C)\rangle &=\delta_{ij}\left(D(z)+D_1(z) +z_4^2 \frac{d D_1(z)}{d z^2}\right) \nonumber \\ &+z_iz_j \frac{dD_1(z)}{dz^2}\;,\\
      \langle O_B (C, -C)\rangle &=\delta_{ij}\left(D(z)+D_1(z) +\vec{z}^2 \frac{d D_1(z)}{d z^2}\right) \nonumber \\ &-z_iz_j \frac{dD_1(z)}{dz^2}\;,\label{gaugeinv}
\end{align}
where $D(z), D_1(z)$ are exponentially decaying functions. Our results, \eqref{electricvortex} and \eqref{bvortex}, at low momenta indicate that the electric correlations are heavily suppressed, whereas the magnetic ones indicate an approximately constant transverse part and a constant longitudinal part with a $k^2$ leading behavior. We have found that the Fourier transform of \eqref{gaugeinv} does not impose relevant restrictions on our parameters, as the low-momentum behavior is very sensitive to minimal changes in position space. Further investigations of this momentum behavior would be welcomed as they could shed light on center-vortex signatures in the infrared properties of the gauge-invariant correlation functions. 

\section{Conclusions}

In this Letter, field correlators were obtained for the first time within a theoretical framework for confinement based on center vortices, which thereby incorporates the asymptotic $N$-ality properties observed in lattice simulations of pure Yang-Mills theory. This was done by means of a wave functional peaked at a center-vortex ensemble. In previous works, it was shown that the nonoriented component causes the Wilson-loop center-element average at asymptotic distances to be dominated by a saddle point or domain wall on the minimal surface bounded by the quark loop \cite{mixed,wavefunctional,3densemble}.
Here, we identified the gauge-invariant field-strength correlator as a natural object to be computed in the center-vortex scenario. Indeed, the wave functional description allows one to trace the probability amplitude of locally Abelian gauge-field variables, which can parametrize the essential elements in the ensemble. The gauge-invariant two-point correlator is constructed as usual by including a pair of holonomies between both points. When they are connected by a single line traversed in both orientations, no linking-number effect arises and no center-element averages are generated. Therefore, no solitonic domain wall is induced. However, the calculation shows the presence of a mass gap generated by the nonoriented component. This component also proved crucial for generating confining flux tubes with $N$-ality and the asymptotic Casimir scaling law.

\section{Acknowledgments}
This work was financed, in part, by the São Paulo Research Foundation (FAPESP), Process  No. 2023/18483-0 (D. R. J.), No. 2024/20896-3 (D. R. J.), No. 2025/15966-5 (B. R. S.), and No. 2018/25225-9
(G. K.); Conselho Nacional de Desenvolvimento Científico e Tecnológico (CNPq) Grants No. 13254/2025-7 (G. K.) and No. 309971/2021-7 (L. E. O.); and Coordenação de Aperfeiçoamento de Pessoal de Nível Superior (CAPES) Grant No. 88887.949835/2024-00 (B. R. S.)

\end{document}